\documentclass[11pt,a4paper]{scrartcl}

\usepackage[utf8]{inputenc}

\usepackage{ILD}

\usepackage[symbol]{footmisc}
\usepackage{feynmf}
\usepackage{multirow}
\usepackage{textpos}

\title{The mini-DST: a high-level LCIO format}

\date{11 May 2021}

\addauthor{Shin-ichi Kawada}{\institute{1}}

\addinstitute{1}{DESY, Notkestra{\ss}e 85, 22607 Hamburg, Germany}

\abstract{
A new LCIO-based data format called mini-DST has been developed, which combines Particle Flow Object (PFO) and event-level information, including the output of the most important high-level reconstruction algorithms.
Originally triggered by Snowmass 2021 studies, the mini-DST is useful for beginners as a starting point of an analysis.
In this note, we discuss the basics and contents of the mini-DST, how to create the mini-DST file from fully simulated MC samples, and the limitations of mini-DST.
\footnote{Talk presented at the International Workshop on Future Linear Colliders (LCWS2021), 15-18 March 2021. C21-03-15.1.}
}

\graphicspath{ {./logos/}{./figures/} }

\begin{document}

\titlepage

\section{Introduction}
\label{sec:intro}
The physics analysis using MC samples is very important to evaluate any kind of physics performance of future experiments.
To make the analysis as realistic as possible, one should use fully simulated MC samples, if available.
Another option is to use the MC samples based on fast simulation, \textit{e.g.} DELPHES~\cite{DELPHES} or SGV (\textit{Simulation \`{a} Grande Vitesse})~\cite{SGV}.
In ILD, most physics analyses are performed with fully simulated MC samples and the events are stored in the DST data format.

However, MC samples normally contain vast amounts of information which will increase the complexity of usage, especially for beginners such as bachelor students, summer students, and newcomers who might have little experience to use such samples.
For example, there are more than 20 collections stored in the DST file which is created with iLCSoft version 02-02 for the ILD-IDR~\cite{ILDIDR}.
The mini-DST project was started to reduce such complexity for beginners.
The idea of the mini-DST is to provide a simple data format and analysis environment without installing the entire iLCSoft~\cite{ilcsoft} as the starting point for the analysis.
This project was originally triggered by The Particle Physics Community Planning Exercise, a.k.a Snowmass 2021~\cite{Snowmass}.

\begin{figure}[h]
    \centering
    \includegraphics[width=0.85\textwidth]{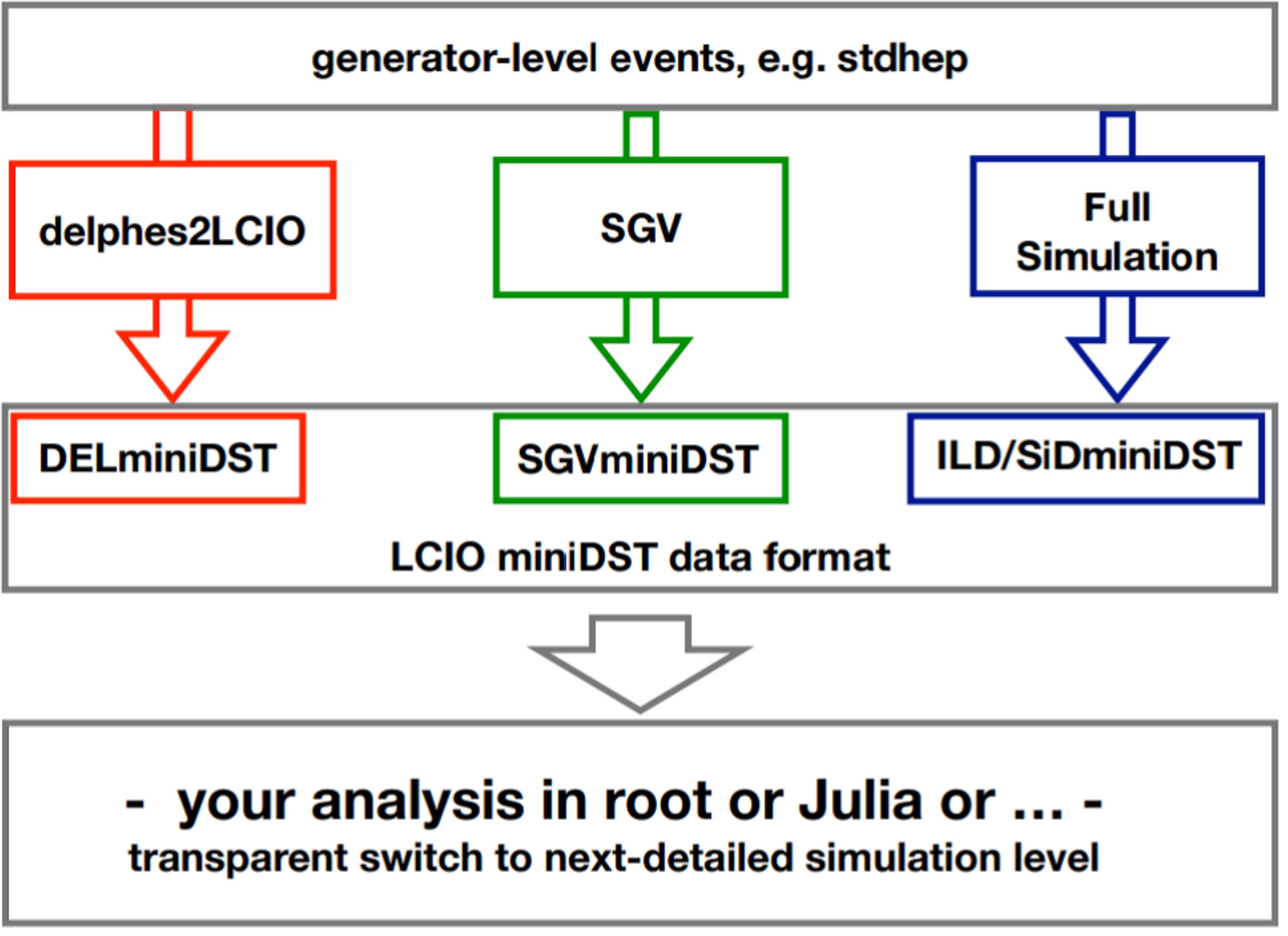}
    \caption{Concept of mini-DST.}
    \label{fig:concept_mini-DST}
\end{figure}

The concept of the mini-DST can be seen in Figure~\ref{fig:concept_mini-DST}.
Since the mini-DST is a data format, it can be applied for both full and fast simulation.
This allows users to use the same analysis code to compare the fast and full simulation.
In this note, we will focus on how to create a mini-DST file from a full simulation.
The details of the DELPHES ILC card are available under~\cite{ILCDelphes} and~\cite{Delphes2lcio}.
The SGV-based mini-DST can be created in a very similar way to a full simulation one.

In this note, we discuss the basics of the mini-DST format, how to make mini-DST files from ILD MC samples, and its limitations.
This document is structured as follows.
We explain the basics and the contents of the mini-DST in Section~\ref{sec:contents}, then we discuss how to make a mini-DST file using a fully simulated MC sample in Section~\ref{sec:howtomake}.
The usage of the mini-DST is summarized with examples in Section~\ref{sec:use}.
Section~\ref{sec:summary} is the summary.

\section{Contents of mini-DST}
\label{sec:contents}
Various types of collections of objects are stored in the DST files.
It includes collections related to vertices, tracks, calorimeter clusters, reconstructed particles, as well as MC truth information.
The mini-DST contains selected collections from the original DST and additional high-level objects.
Table~\ref{tab:contents} shows the contents of the mini-DST and an explanation of each element.
The mini-DST in particular contains high-level objects such as isolated objects and jets which are not present on the DST.
These high-level objects are very useful for most analyzes.
Their definition and the algorithms used to create them are specified in Section~\ref{sec:howtomake}.
The usage of mini-DST will be described in Section~\ref{sec:use}.
Note that some collections are not available in DELPHES-mini-DST, and the name of collections are might be different for each type of mini-DST.
Table~\ref{tab:contents} shows the case of mini-DST of full simulation.

\begin{table}[h]
    \centering
    \caption{Contents of mini-DST.
    Details of the various classes can be found at~\cite{LCIO}.}
    \begin{tabular}{ccl} \hline
        Collection name & Collection type & Explanation \\ \hline
        PandoraPFOs & ReconstructedParticle & particle flow objects plus event shape variables \\
        PFOsminusoverlay & ReconstructedParticle & PandoraPFOs minus $\gamma \gamma \to$ low $P_t$ hadrons background \\
        BCalPFOs & ReconstructedParticle & \begin{tabular}{l} particle flow objects from the most \\ forward calorimeter \end{tabular} \\
        PrimaryVertex & Vertex & primary vertex \\
        PrimaryVertex\verb|_|RP & ReconstructedParticle & \begin{tabular}{l} "reconstructed particle" representing the \\ primary vertex \end{tabular} \\
        IsolatedElectrons & ReconstructedParticle & isolated electrons \\
        IsolatedMuons & ReconstructedParticle & isolated muons \\
        IsolatedTaus & ReconstructedParticle & isolated taus \\
        IsolatedPhotons & ReconstructedParticle & isolated photons \\
        RefinedNJets & ReconstructedParticle & \begin{tabular}{l} PandoraPFOs minus isolated objects forced into \\ N jets (N = 2, 3, 4, $\dots$ up to 10, Durham clustering, \\ plus flavor tag) \end{tabular} \\
        MCParticlesSkimmed & MCParticle & MC truth information \\
        MCTruthRecoLink & LCRelation & links from MCParticlesSkimmed to PandoraPFOs \\
        RecoTruthMCLink & LCRelation & links from PandoraPFOs to MCParticlesSkimmed \\ \hline
    \end{tabular}
    \label{tab:contents}
\end{table}

\section{Creation of mini-DST files}
\label{sec:howtomake}
In this section, we will discuss the technical details of how to make a mini-DST file using a DST file produced with ILD.
The processors necessary to create mini-DST are contained in iLCSoft.
The parameters of each processor are used as the default value unless stated.

\subsection{First step}
\label{sec:first}
Before producing mini-DSTs, one should prepare the input DST file and initialize iLCSoft in the terminal.
This note are based on version 02-02-01 of iLCSoft.
One needs to run the InitializeDD4hep~\cite{MarlinDD4hep} processor to initialize DD4hep, otherwise, some functionalities do not work and jobs will fail.

In this note, we will only discuss using MC samples produced in the context of the ILD-IDR and later produced MC samples.
If one wants to create mini-DST from DBD~\cite{ILDDBD} samples, the beam polarization information is not stored in the event record but needs to be extracted from the file name.
For this, we have developed a simple processor to add the beam polarization in the event header based on the input DST file name for DBD samples.
It is available at the author's personal GitHub branch~\cite{AddPol}.
However, it is recommended to use MC samples that are recently produced.

\subsection{$\gamma \gamma \to$ low $P_t$ hadrons overlay}
\label{sec:overlay}
For lower center-of-mass energies like 250~GeV and 350~GeV, the contribution of $\gamma \gamma \to$ low $P_t$ hadrons background is small and thus no treatment is applied to remove this background from the mini-DST.

For higher center-of-mass energies, 500~GeV and beyond, the contribution of $\gamma \gamma \to$ low $P_t$ hadrons is not negligible anymore.
Therefore, we use the FastJetProcessor~\cite{MarlinFastJet} to remove this background.
The input collection is PandoraPFOs and the output collection is PFOsminusoverlayJets.
For the 500~GeV ILD-IDR MC samples, we perform the $e^+ e^-$-generalized $k_T$ clustering with $p = 1.0$ and the jet radius of $R = 3.0$ requiring number of jets $N = 2$.
Details of the parametrization are discussed in Appendix~\ref{app:param}.

Note that the output from FastJetProcessor is a jet collection, not a collection of individual reconstructed particles.
We use JetPFOsProcessor~\cite{OverlayRemoval} to unpack jets to PFOs.
The input collection is PFOsminusoverlayJets and the output collection is PFOsminusoverlay.

\begin{figure}[h]
    \centering
    \includegraphics[width=0.6\textwidth]{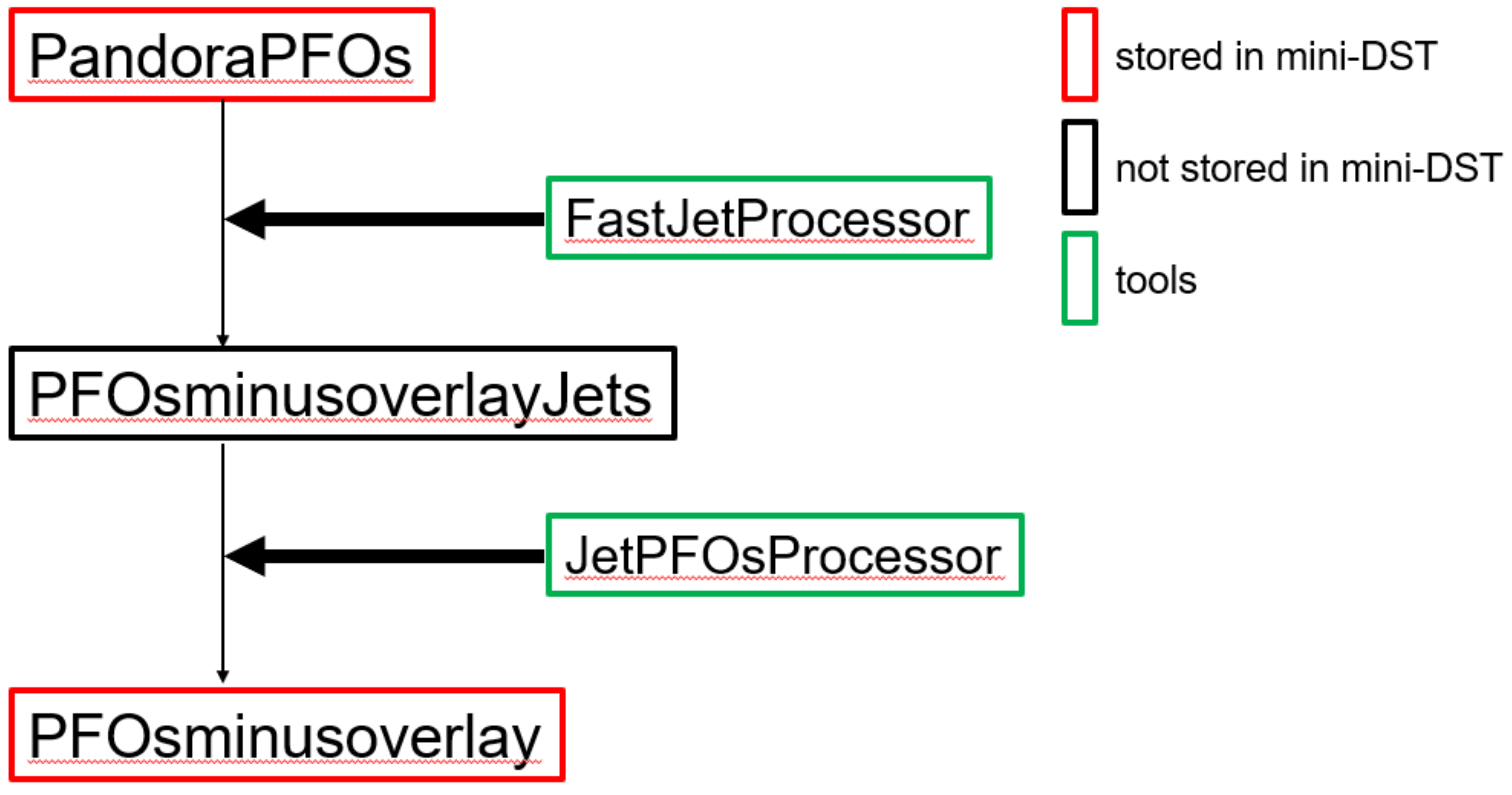}
    \caption{Flow chart of $\gamma \gamma \to$ low $P_t$ hadrons overlay.}
    \label{fig:flow1}
\end{figure}

\subsection{Event shape variables}
\label{sec:EventShape}
We use the Marlin processors ThrustReconstruction, Sphere, and Fox in order to calculate the thrust, sphericity, and Fox-Wolfram moments, all stored in the same repository~\cite{EventShapes}.
The input collection for these processors is PandoraPFOs (or PFOsminusoverlay).
The calculated results are stored in the header of the PandoraPFOs (or PFOsminusoverlay) collection.

Note that these event shape variables are calculated using all particles in an event (except $\gamma \gamma \to$ hadrons which are removed in Section~\ref{sec:overlay}).
If one wants to calculate, for example, the thrust value without including the isolated objects in an event, one needs to prepare a corresponding input collection and run the processor again.

\subsection{Isolated muons and electrons}
\label{sec:IsolatedMuonsElectrons}
In the remaining steps, we will tag specific particles as isolated objects, starting from muons and electrons.
We start by finding isolated muons using the IsolatedLeptonTaggingProcessor~\cite{IsolatedLeptonTagging}.
The input collection is PandoraPFOs (or PFOsminusoverlay) and the output collection is PFOsminusmu.
The IsolatedLeptonTaggingProcessor is based on a multivariate discriminator.
The location of the weight files from the training process needs to be specified in \verb|DirOfISOMuonWeights|.
The parameter \verb|IsSelectingOneIsoLep| should be set as \verb|false| to capture all isolated muons.
The parameter \verb|CutOnTheISOElectronMVA| needs to be set to a value above 1 to not capture any isolated electrons.
The selected particles will be stored as the IsolatedMuons collection.

After removing the identified isolated muons, we will capture the isolated electrons.
We again use IsolatedLeptonTaggingProcessor.
The input collection is PFOsminusmu and the output collection is PFOsminuse.
As we did in isolated muon finding, the parameter \verb|IsSelectingOneIsoLep| should be set as \verb|false| to capture all isolated electrons, and \verb|CutOnTheISOMuonMVA| needs to be set to a value above 1 to not to capture any isolated muons.
The selected particles will be stored as IsolatedElectrons collection.

Note that the isolated electrons and muons from tau leptons
will be included in IsolatedMuons and IsolatedElectrons.

\begin{figure}[h]
    \centering
    \includegraphics[width=0.6\textwidth]{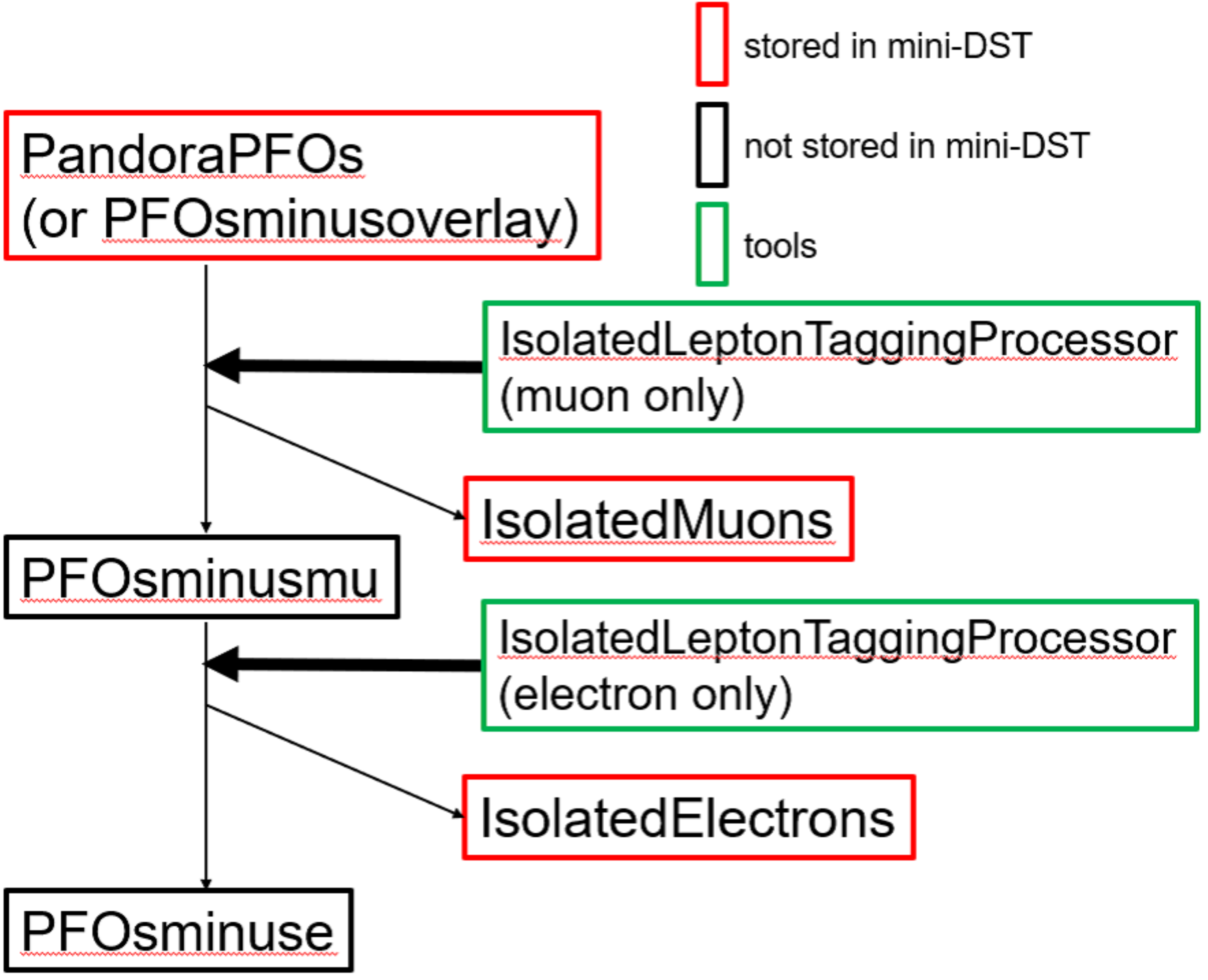}
    \caption{Flow chart of isolated muons and electrons.}
    \label{fig:flow2}
\end{figure}

\subsection{Isolated taus}
\label{sec:IsolatedTaus}
Finding isolated taus is performed using the processor TaJetClustering~\cite{TauFinder}.
This processor was originally developed to identify tau leptons under jet background.
The input collection is PFOsminuse and the output collection is PFOsminustau.
The selected particles will be stored as IsolatedTaus.

Note that this processor adopts a simple cut-based technique, leaving room for future improvement by more sophisticated techniques.

\begin{figure}[h]
    \centering
    \includegraphics[width=0.5\textwidth]{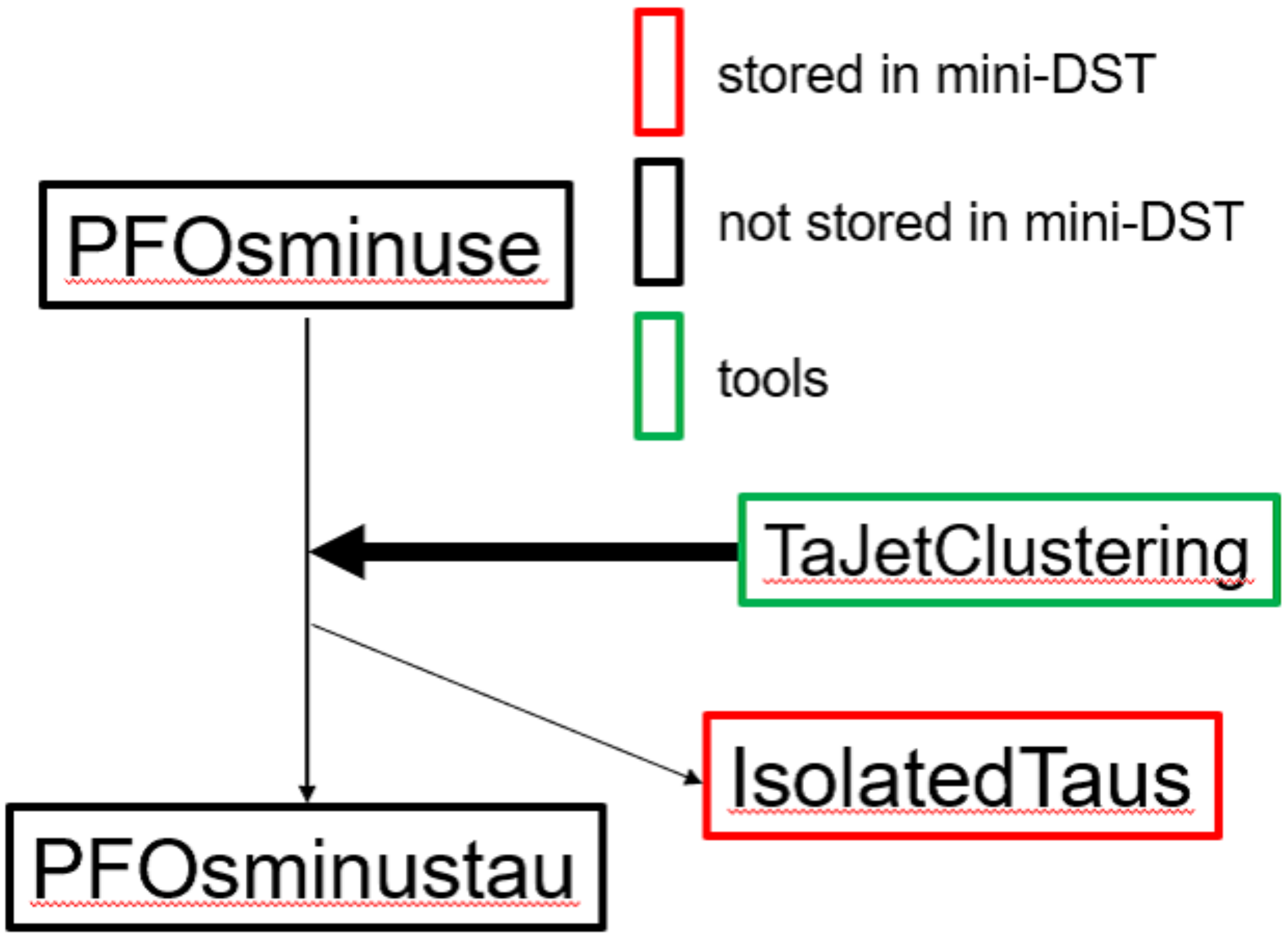}
    \caption{Flow chart of isolated taus.}
    \label{fig:flow3}
\end{figure}

\subsection{Isolated photons}
\label{sec:IsolatedPhotons}
At this point we will attempt to find isolated photons.
For this, we use the IsolatedPhotonTaggingProcessor~\cite{IsolatedLeptonTagging} which is located in the same repository as the IsolatedLeptonTaggingProcessor.
The input collection is PFOsminustau and the output collection is PFOsminusphoton.
The selected particles will be stored as IsolatedPhotons.

Like the TaJetClustering processor, this processor is also using a cut-based technique at present, so also here there is room for future improvement


\begin{figure}[h]
    \centering
    \includegraphics[width=0.6\textwidth]{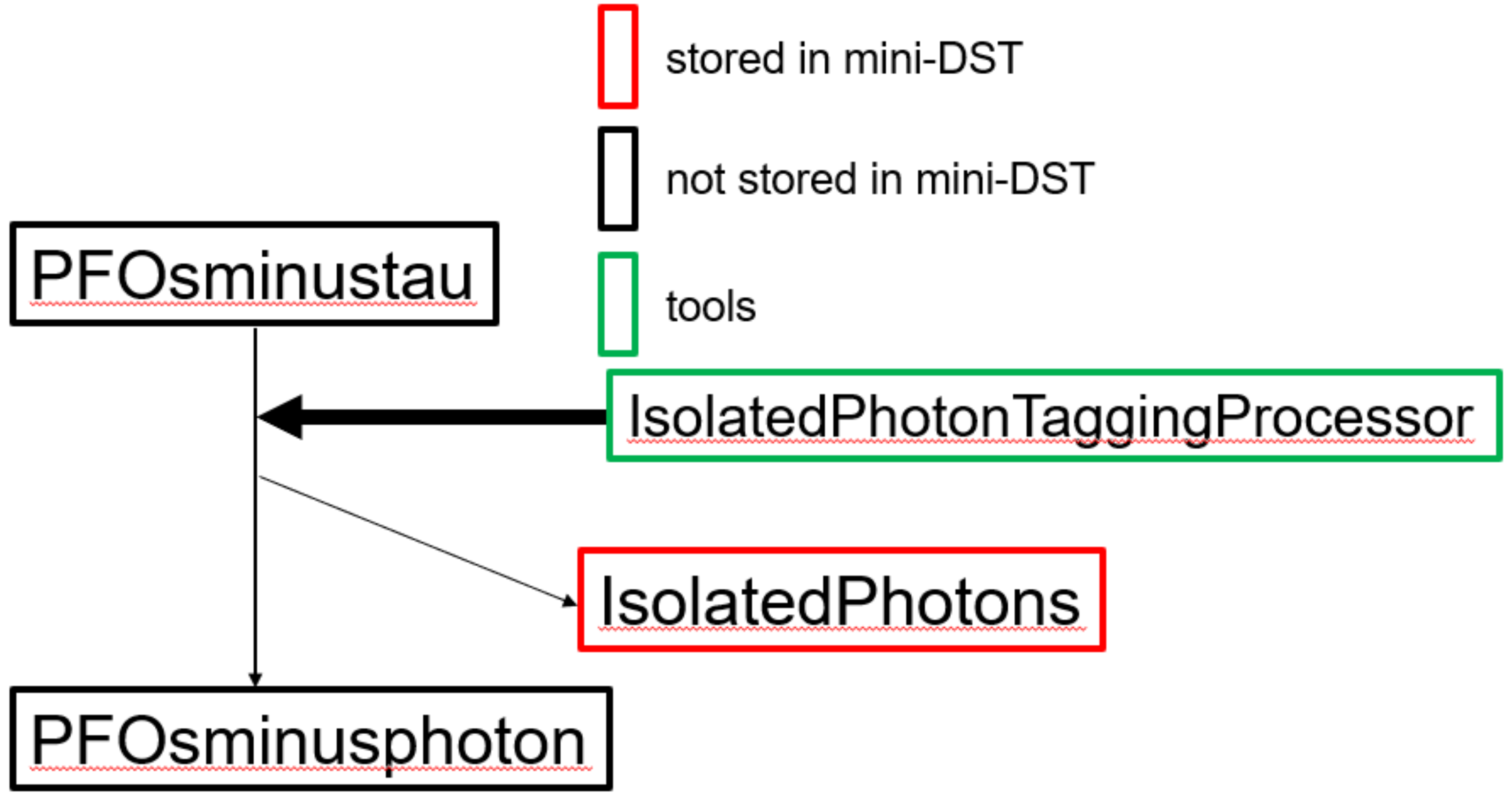}
    \caption{Flow chart of isolated photons.}
    \label{fig:flow4}
\end{figure}

\subsection{Jet clustering and flavor tagging}
\label{sec:LCFIPlus}
Finally, we perform jet clustering and flavor tagging using the LcfiplusProcessor~\cite{LCFIPlus}.
The input collection is PFOsminusphoton and the output collection is RefinedNJets.
The N corresponds to the number of jets in the jet clustering.
The LcfiplusProcessor will produce multiple collections, but only the selected collections are stored.

We perform jet clustering requiring a fixed number of jets.
At $\sqrt{s} =$ 250 GeV, we create RefinedNJets collections with N of 2, 3, 4, 5, and 6.
At higher center-of-mass energies, the value of \verb|N| is increased to 10.

Finally, the ErrorFlow~\cite{ErrorFlow} is applied to calculate the covariance matrix of the jets.
The ErrorFlow processor directly modifies its input collection, RefineNJets.
It also creates a dummy output collection, which is dropped from the mini-DST.

\begin{figure}[h]
    \centering
    \includegraphics[width=0.7\textwidth]{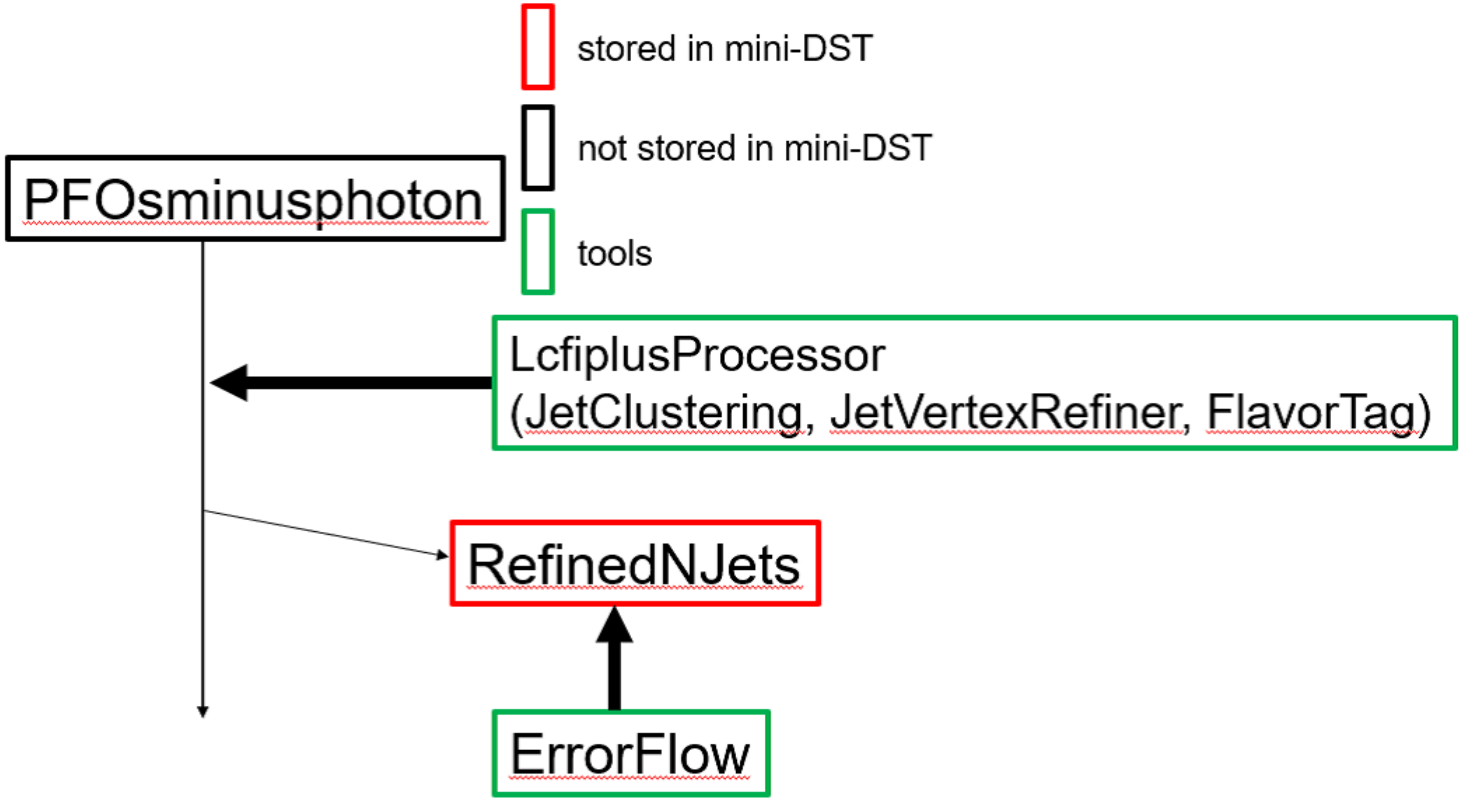}
    \caption{Flow chart of jet clustering and flavor tagging.}
    \label{fig:flow5}
\end{figure}

\subsection{Final steps}
\label{sec:final}
After adding all the collections described in Section~\ref{sec:LCFIPlus}, the final mini-DST file is written using the LCIOOutputProcessor~\cite{LCIOOutput}.
We can control which collections should be included and/or removed with this processor.
We must set the name of the output file as the parameter.
Most of the collections are removed and only the selected collections are stored as listed in Table~\ref{tab:contents}.

\begin{figure}[h]
    \centering
    \includegraphics[width=0.95\textwidth]{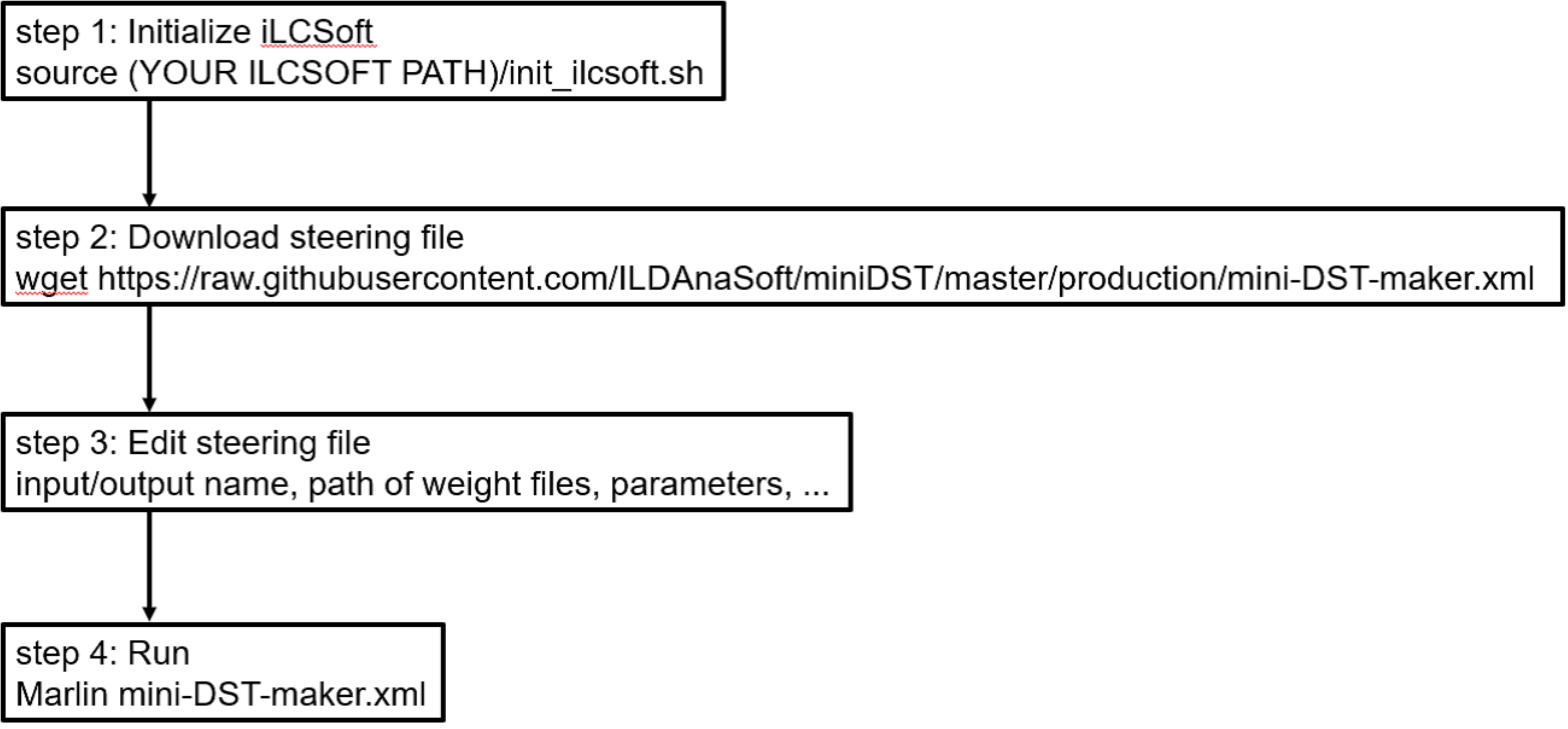}
    \caption{Flow chart of mini-DST production.}
    \label{fig:flow6}
\end{figure}

The flow of mini-DST production is graphically summarized in Figure~\ref{fig:flow6}.
For the production of a mini-DST file, one needs to initialize iLCSoft and to set parameters of processors (input file, name of output file, path of weight files of IsolatedLeptonTagging and LcfiplusProcessor, and more if need).
Assuming that the steering file is called \verb|mini-DST-maker.xml|, the command \verb|Marlin mini-DST-maker.xml| on the command-line will launch the production.
If one needs mass production of mini-DST files, one should consider using the computing system with many CPUs.
The production of SGV-based mini-DST is identical, only replace the input DST file with an SGV file.

Note that the production of mini-DST files may take hours when the number of events of the input file is large and/or using high multiplicity events (like $e^+ e^- \to t\overline{t}$ events).
This is mainly due to the long-running time of the LcfiplusProcessor.

\section{Usage of mini-DST}
\label{sec:use}
In this section, we will discuss how to use the mini-DST with an example and simple macro.
The limitations of the mini-DST will also be discussed.

\subsection{Example: recoil mass study at ILC250}
\label{sec:ILC250}
In this note, we take the simplest example from the tutorial which is available at ~\cite{tutorial}.
A few more examples are available at~\cite{tutorial_slide} and~\cite{usemini-DST}.

The ILC will start its operation with the center-of-mass energy of 250~GeV.
The recoil mass study is one of the most important research topics at the ILC250.
In this example, we will discuss the process of $e^+ e^- \to ZH \to (\mu ^+ \mu ^-)(jj)$.
Since the initial 4-momentum is known, the Higgs mass can be calculated with the following formula once the 4-momentum of muons from $Z$ boson are measured;
\begin{equation}
    M_H^2 = \left( p_{\mathrm{initial}} - \left( p_{\mu ^+} + p_{\mu ^-} \right) \right)^2,
\end{equation}
where $p$ is the 4-momentum.

\begin{figure}[h]
\begin{tabular}{ccc}
\begin{minipage}{0.305\textwidth}
    \centering
    \includegraphics[width=\textwidth]{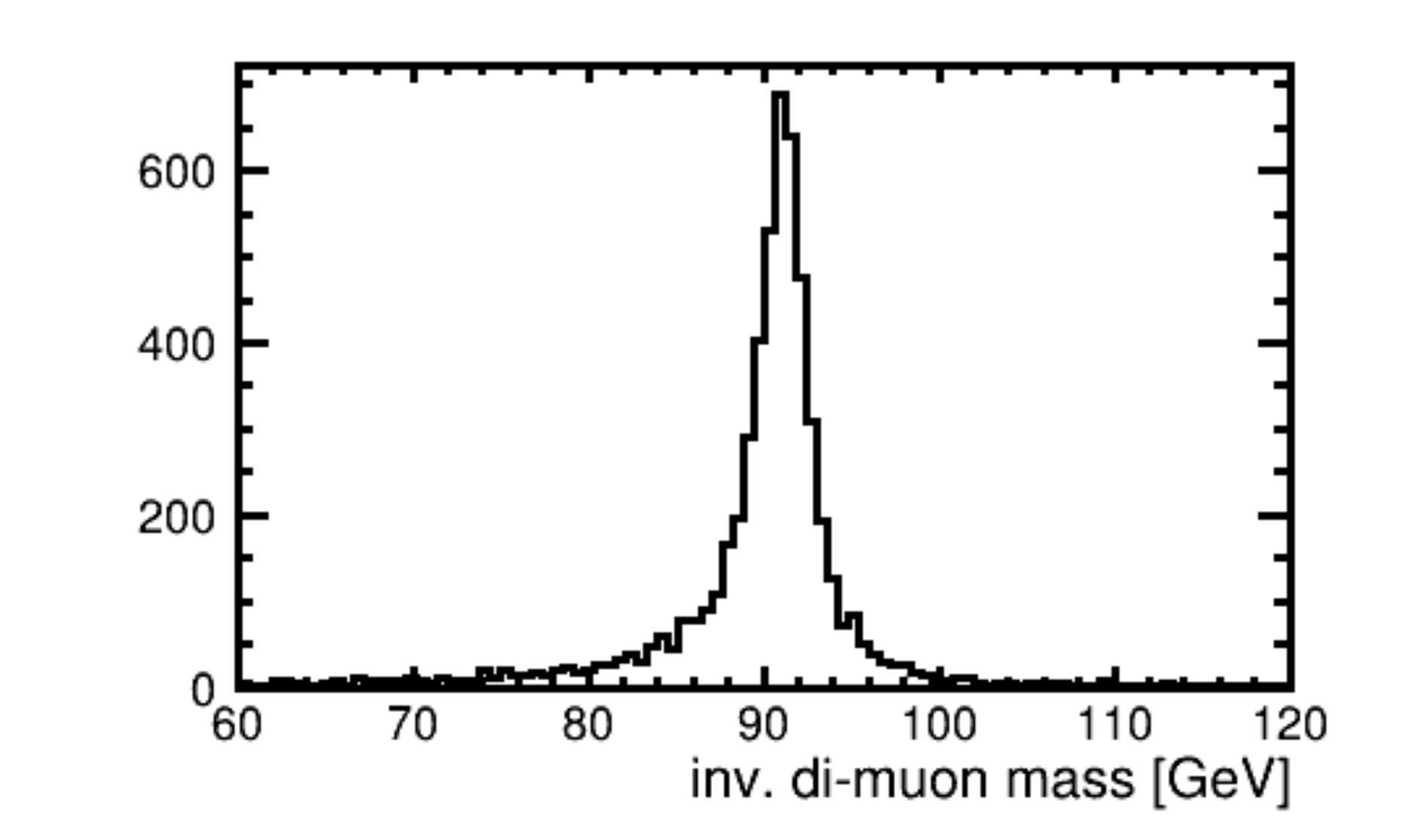}
    \subcaption{$M_{\mu ^+ \mu ^-}$.}
    \label{sub:a}
\end{minipage} &
\begin{minipage}{0.305\textwidth}
    \centering
    \includegraphics[width=\textwidth]{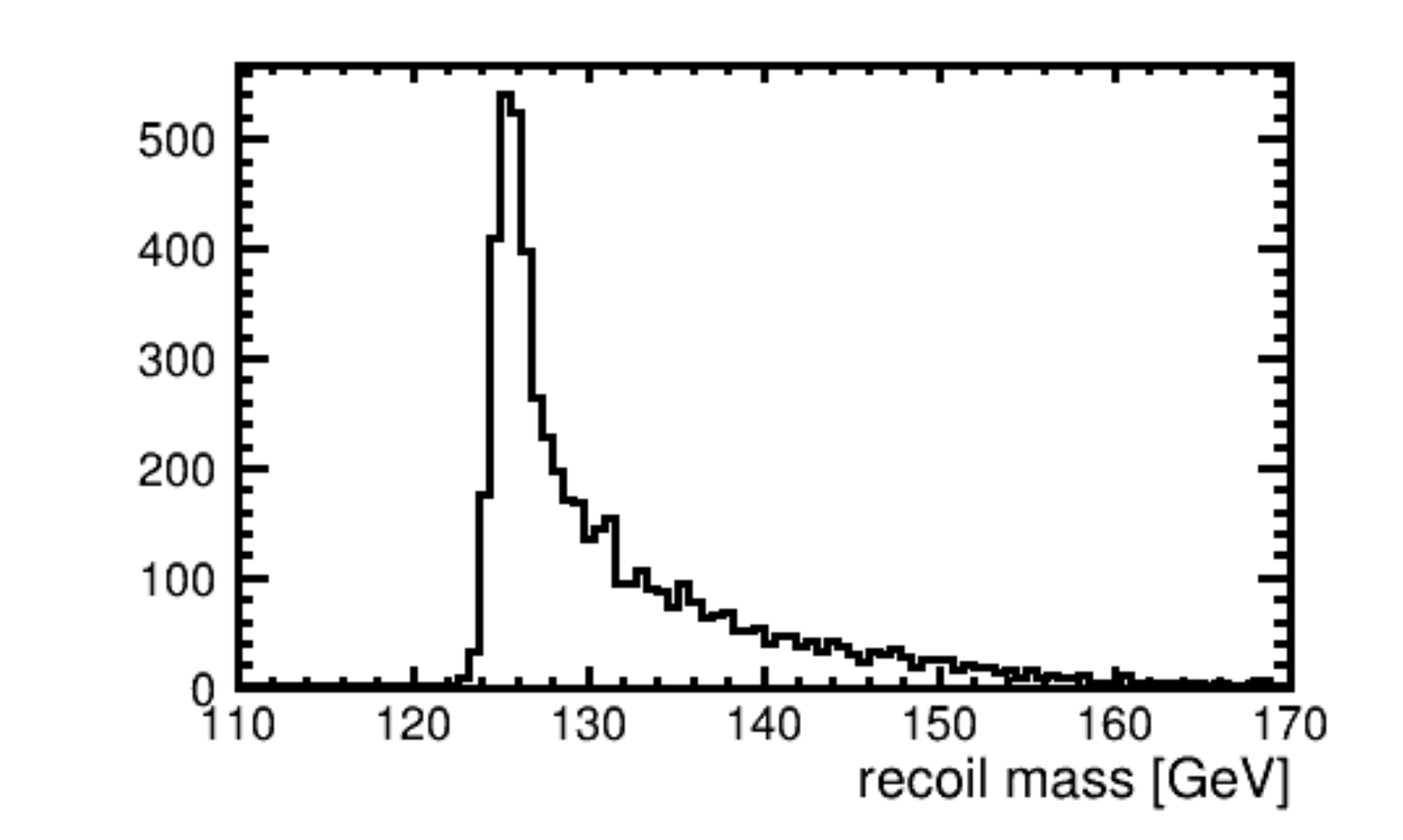}
    \subcaption{$M_{\mathrm{recoil}}$.}
    \label{sub:b}
\end{minipage} &
\begin{minipage}{0.305\textwidth}
    \centering
    \includegraphics[width=\textwidth]{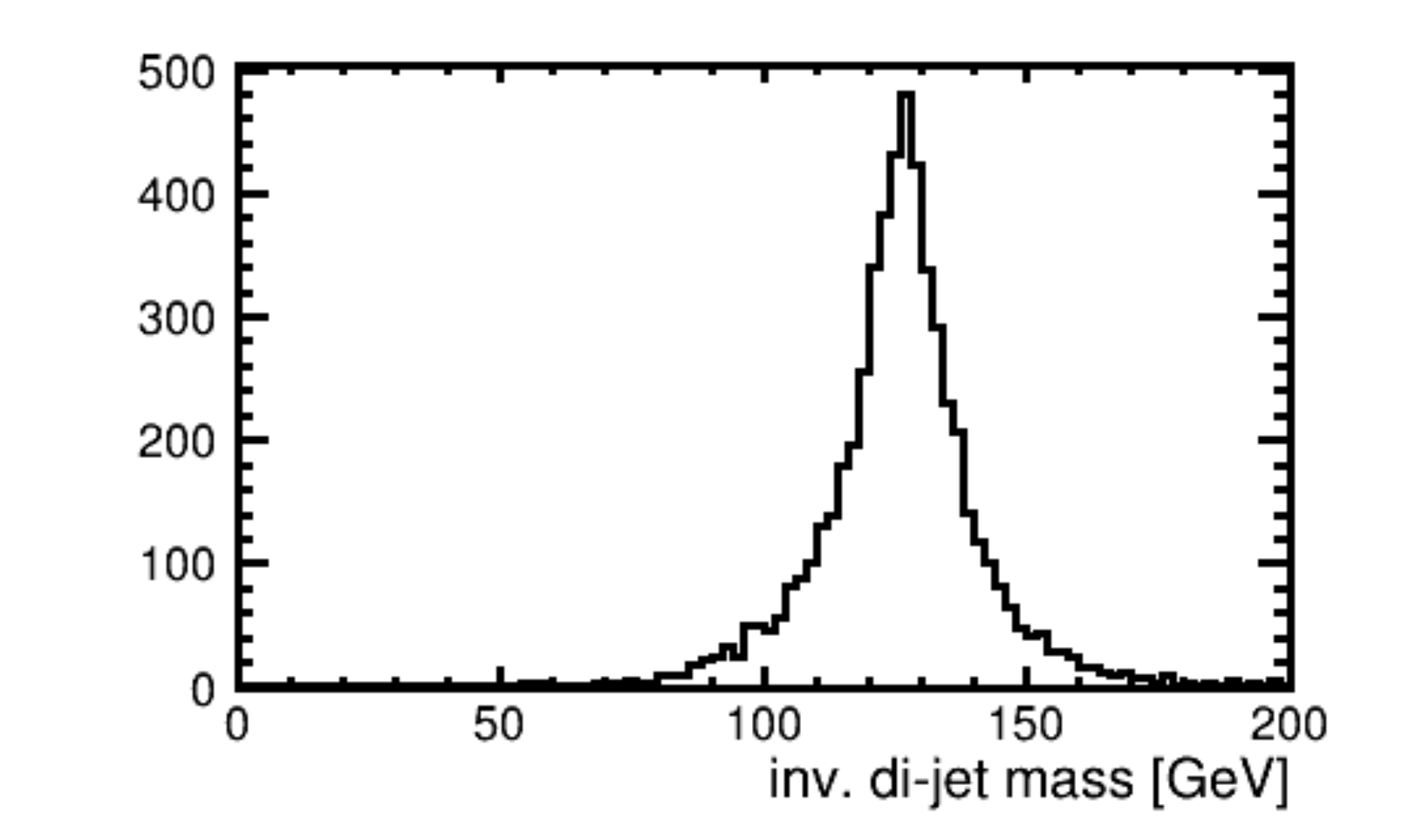}
    \subcaption{$M_{jj}$.}
    \label{sub:c}
\end{minipage}
\end{tabular}
\caption{Example histograms.}
\label{fig:example}
\end{figure}

One only needs to define the ROOT environment, the LCIO libraries, the input mini-DST file, and a macro named higgs\verb|_|recoil.C which will discuss now.
Include the following lines (Code~\ref{code01}) in .rootlogon.C file and execute one command (Code~\ref{code02}) on the command-line to use LCIO.
\begin{lstlisting}[caption=.rootlogon.C., label=code01]
{
gInterpreter->AddIncludePath("$LCIO");
gSystem->Load("${LCIO}/lib/liblcio.so");
gSystem->Load("${LCIO}/lib/liblcioDict.so");
}
\end{lstlisting}
\begin{lstlisting}[caption=Enables to use LCIO., label=code02]
export = LD_LIBRARY_PATH=$LCIO/lib:$LD_LIBRARY_PATH
\end{lstlisting}
Then start a ROOT session and type \verb|.x higgs_recoil.C("FILENAME", "OUTPUTNAME")|.
The histograms shown in Figure~\ref{fig:example} will be produced.
The parameter \verb|FILENAME| is the name of the mini-DST file.
Figure~\ref{sub:a} shows the muon pair invariant mass $M_{\mu ^+ \mu ^-}$, \ref{sub:b} shows the recoil mass against muon pair $M_{\mathrm{recoil}}$, and~\ref{sub:c} shows dijet mass $M_{jj}$.

The core part of higgs\verb|_|recoil.C is shown in Code~\ref{code03}.
In this code, the following steps are performed.
\begin{itemize}
    \item read the collections which one wants to read,
    \item take 4-momentum of muons and calculate its invariant mass ($Z \to \mu ^+ \mu ^-$),
    \item calculate recoil mass considering beam crossing angle effect,
    \item take 4-momentum of jets and calculate its invariant mass ($H \to jj$),
    \item write histograms of variables one wants to see.
\end{itemize}
\begin{lstlisting}[caption=The core part of the macro., label=code03]
while( ( evt = lcReader->readNextEvent() ) != 0  && nEvents++ < maxEvt ){
   LCIterator<ReconstructedParticle> jets( evt, "Refined2Jets" );
   LCIterator<ReconstructedParticle> muons( evt, "IsolatedMuons" );

   //only use events with 2 jets and 2 muons
   if( jets.size() != 2 ) continue;
   if( muons.size() != 2 ) continue;

   //the dimuon mass
   auto mu1 = muons.next(); 
   auto mu2 = muons.next();    
   hmuonmass->Fill( inv_mass( mu1, mu2 ) );

   //the recoil mass
   const auto& vm1 = v4(mu1);
   const auto& vm2 = v4(mu2);
   double pxinitial = 0.;
   double Einitial = 250.;
   // in full sim & SGV, correct for crossing angle
   if (!isDelphes) {  
     pxinitial = Einitial*0.007; 
     Einitial = 2.*std::sqrt(std::pow(Einitial/2.,2) + std::pow(pxinitial,2));
   }
   TLorentzVector ecms(pxinitial,0.,0.,Einitial);
   TLorentzVector recoil = ecms - ( vm1 + vm2 );
   hrecoilmass->Fill( recoil.M() );
   
   //the dijet mass
   auto j1 = jets.next();
   auto j2 = jets.next();
   hjetmass->Fill( inv_mass( j1, j2 ) );
}
\end{lstlisting}

\subsection{Limitations of the mini-DST}
\label{sec:limit}
The original philosophy of the mini-DST is to provide a simple data format as a starting point of an analysis.
Most of the processors to create high-level objects are run with default parameter values as discussed in Section~\ref{sec:howtomake}.
Thus, the mini-DST has some limitations.

The first limitation is coming from the simplicity of collections in the mini-DST.
As shown in Table~\ref{tab:contents}, most of the collections in the mini-DST are of type ReconstructedParticle.
This means that some collections originally stored in DST files are not stored in mini-DST files.
For example, it is impossible to access the collections related to tracks and clusters using the mini-DST because these are not stored.
If such collections are needed, one should consider switching from mini-DST to DST or produce special samples.

The second limitation is coming from the default choices of the steering parameters for most of the processors, which provide a good starting point, but might not be optimal for every analysis.

\section{Summary and future}
\label{sec:summary}
The mini-DST data format has been developed.
The philosophy of the mini-DST is to provide an easy starting point for analyses, especially for beginners.
In this note, we discuss the basics of the mini-DST, how to create a mini-DST file from the original DST file, and the limitations of the mini-DST format.

The mini-DST is still developing.
In the future, we are planning to add an \verb|EventSummary| feature which makes the analysis even faster~\cite{EventSummary}.
In the \verb|EventSummary|, the number of isolated objects, $Y_{\mathrm{cut}}$-value of jet clustering, event shapes, missing 4-momentum, and other event-level quantities will be stored.
One can then easily choose the event type which one wants to see (\textit{e.g.} events with two isolated muons) via the \verb|EventSummary|.

\section*{Acknowledgements}
We would like to thank the LCC generator working group and the ILD software working group for providing the simulation and reconstruction tools and producing the Monte Carlo samples used in this study.
This work has benefited from computing services provided by the ILC Virtual Organization, supported by the national resource providers of the EGI Federation and the Open Science GRID, and from computing services provided by the German National Analysis Facility (NAF)~\cite{NAF}.
We thankfully acknowledge the support by the the Deutsche Forschungsgemeinschaft (DFG, German Research Foundation) under Germany’s Excellence Strategy EXC 2121 “Quantum Universe” 390833306.

\printbibliography{}

\appendix
\section{Parametrization in FastJetProcessor}
\label{app:param}
\begin{figure}[h]
    \centering
    \includegraphics[width=0.9\textwidth]{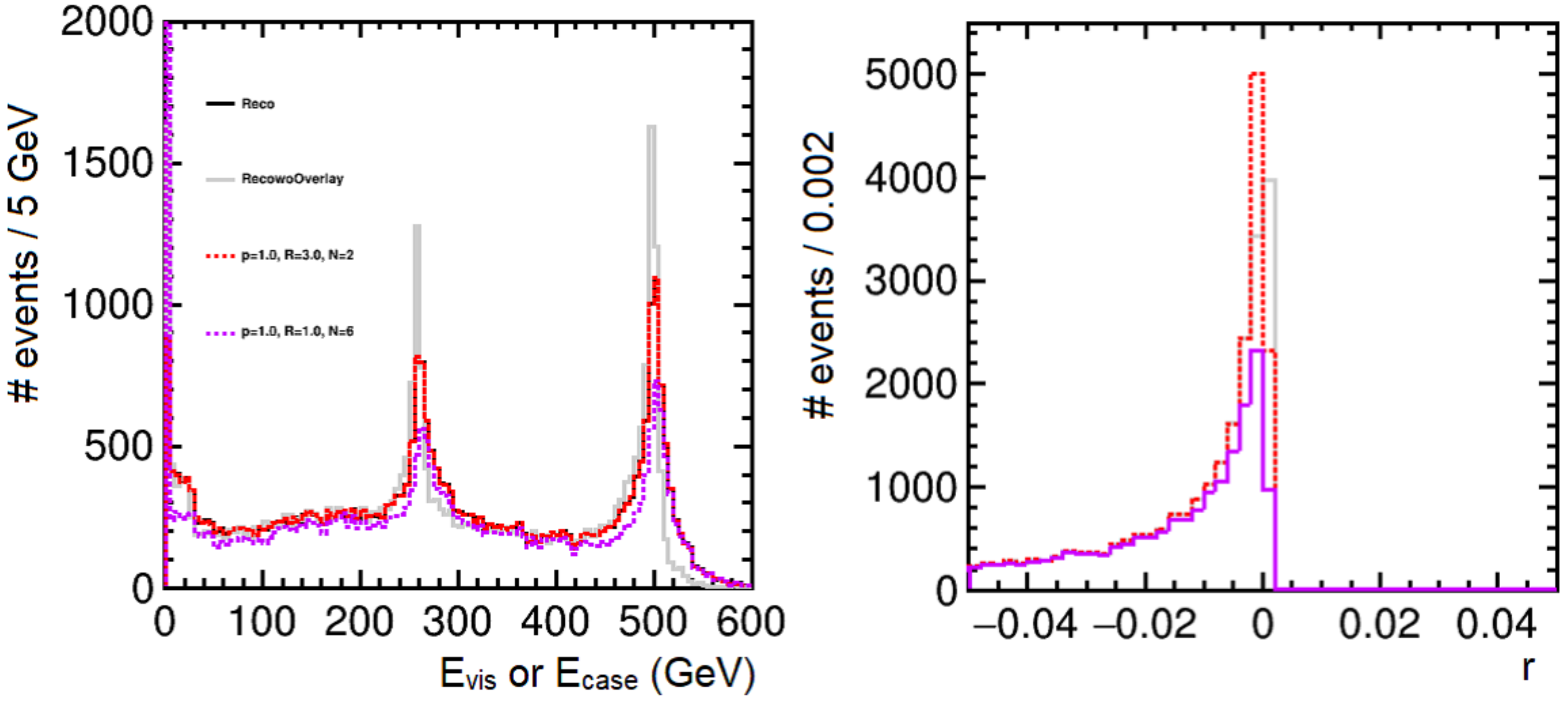}
    \caption{Comparison between different parameter choices for the FastJetProcessor.
    The ash histograms show the reconstructed particles minus the $\gamma \gamma \to$ hadrons using MC truth information, the red histograms show the reconstructed particles with $p = 1.0$, $R = 3.0$, $N = 2$, and purple histograms show the reconstructed particle with $p = 1.0$, $R = 1.0$, $N = 6$.
    Left: visible energy of an event $E_{\mathrm{vis}}$ and $E_{\mathrm{case}}$.
    Right: the residual $r$.}
    \label{fig:comparison}
\end{figure}

In this appendix, we corroborate the choice of parameters for the FastJetProcessor for the removal of $\gamma \gamma \to$ hadrons background contribution.
Figure~\ref{fig:comparison} shows the comparison with different parameter choices for the FastJetProcessor using the 500~GeV IDR \verb|2f_l| sample.
The left plot shows the visible energy of an event subtracting the contribution from $\gamma \gamma \to$ hadrons overlay using MC truth information $E_{\mathrm{vis}}$ in ash color, and the same visible energy for two different parameter choices for the FastJetProcessor $E_{\mathrm{case}}$ in red/purple color.
The red color is the result of FastJetProcessor with $p = 1.0$, $R = 3.0$, $N = 2$, and purple color shows the result with the parametrization of $p = 1.0$, $R = 1.0$, $N = 6$.
The right plot shows the residual $r \equiv (E_{\mathrm{vis}} - E_{\mathrm{case}}) / E_{\mathrm{vis}} = 1 - E_{\mathrm{case}} / E_{\mathrm{vis}}$.
We studied various parameter combinations and concluded that the parametrization of $p = 1.0$, $R = 3.0$, $N = 2$ in FastJetProcessor will give somewhat reasonable and stable results for the $\gamma \gamma \to$ hadrons background removal.

\end{document}